\documentclass[a4paper,aps,pra,preprintnumbers]{revtex4}
\usepackage[utf8]{inputenc}
\usepackage[english]{babel}
\usepackage[T1]{fontenc}
\usepackage{amssymb,amsfonts,amsmath,mathtext,enumerate,float,dsfont}
\usepackage{graphics,graphicx,epsfig,epstopdf}
\usepackage{caption}
\usepackage{cmap}
\usepackage{multirow}
\usepackage{indentfirst}
\usepackage[usenames]{color}
\usepackage{amsthm}

\begin{document}
\title{Generation of Squeezed Fock States by Particle-Number Measurements on Multimode Gaussian States}

\author{S. B. Korolev, A. A. Silin} 
\affiliation{St.Petersburg State University, Universitetskaya nab. 7/9, St.Petersburg, 199034, Russia}

\date{\today}

\begin{abstract}
We investigate the generation of squeezed Fock states (SFSs) via particle-number measurements in the modes of multimode Gaussian states. We identify a universal class of $N$-mode Gaussian states for which measuring $N-1$ modes results in the generation of SFSs. The key feature of these states is that the generated SFSs depend only on the total number of detected particles and are independent of their distribution among the detectors. Based on the general form of the wave functions of multimode Gaussian states, we propose a universal scheme for SFS generation. For this scheme, we evaluate the probability of SFS generation and analyze the robustness of the process against imperfections in particle-number-resolving detectors. In addition, we compare the universal scheme with a nonuniversal scheme, in which the generation of SFSs depends on a specific distribution of particle numbers across the detectors. We demonstrate that the universal scheme provides a higher probability of SFS generation, at the cost of increased experimental resources.
\end{abstract}

\maketitle

\section{Introduction}
Modern quantum information science addresses several fundamental questions concerning non-Gaussian states, including their classification \cite{Genoni2007,Genoni2013,Hughes2014,Kenfack_2004,takagi2018convex,Chabaud2020,zhang2020quantifying,Lukas2022,Chabaud2025},  regimes of applicability \cite{Zhuang2018,Walschaers2021}, and practical methods for their generation \cite{Sokolov2020,Asavanant2021,Masalaeva2022,Podoshvedov_2023,bashmakova2023effect}. The importance of these questions is rooted in the central role played by non-Gaussian states in quantum information processing. In particular, non-Gaussian states constitute a necessary resource for universal quantum computation \cite{Braunstein_2005,Lloyd_1999} and exhibit properties that are essential for quantum metrology \cite{Hou2019,Munoz2021}, quantum teleportation \cite{Opatrny2000,Zinatullin2023,Zinatullin_2022,Zinatullin2021}, and quantum cryptography \cite{Lee2019,Guo2019}. Moreover, their nonclassical statistics and structural asymmetries enable applications in quantum error-correction codes \cite{Ralph_2003,hastrup2022all,Bergmann2016,Mirrahimi_2014,schlegel2022quantum,Gottesman2001,Binomial_state}. 

Despite their significance, the practical use of many prominent non-Gaussian states -- such as Schrödinger cat states, Gottesman–Kitaev–Preskill states, and binomial states -- remains limited due to the lack of methods for their accurate generation \cite{Glancy2006,Vasconcelos2010,Etesse2015,Wang2022,He2023,Konno2024,Kudra2022,Sychev2017,Ourjoumtsev_cat2007,Polzik2006,Huang2015,Ulanov2016,Gerrits2010,Takahashi2008,Thekkadath2020,Baeva2023}. This motivates particular interest in non-Gaussian states that can be accurately (with unit fidelity) generated experimentally. 

One important example of such states is the squeezed Fock state (SFS). SFSs form a well-studied class of non-Gaussian states \cite{Marian1991,Kienzler2017,Olivares2006,olivares2005squeezed,Kral1990,NIETO1997135,Kim1989,Xu2015} with applications in quantum error-correction codes \cite{Korolev2024_eror_cor,Bashmakova2025}, quantum metrology \cite{DellAnno,Zhang2019}, and for generating other non-Gaussian states \cite{Baeva2023b,Winnel2024}. Exact generation of SFSs has been demonstrated in schemes based on particle-number measurements performed on one mode of a two-mode Gaussian state obtained by entangling squeezed vacuum states \cite{Korolev2024}. Two main constraints limit the SFS generation procedure. First, due to the difficulty of preparing oscillators in strongly squeezed states \cite{Vahlbruch2016}, it is not feasible to produce SFSs with large squeezing parameters. Second, it remains challenging to realize detectors capable of reliably resolving large particle numbers.

To address these limitations, one may consider transitioning from a two-mode generation scheme to a multimode one, in which a multimode Gaussian state is measured by multiple particle-number-resolving detectors (PNRDs). Under resource-limited conditions, increasing the number of input modes is equivalent to increasing the total energy of the system, which, in principle, may lead to a larger squeezing parameter of the output state. Likewise, increasing the number of PNRDs should enable the measurement of higher particle numbers.

In this work, we investigate the generation of SFSs in schemes based on particle-number measurements performed on the modes of a multimode Gaussian state. We determine the scheme parameters, as well as the squeezing degree of the oscillators used, for which measuring an arbitrary particle number guarantees the generation of an SFS. We analyze the success probabilities of SFS generation and identify the associated experimental constraints. In addition, we examine the influence of detector imperfections on the generated SFS. Finally, we compare the proposed scheme with alternative approaches to SFS generation.

\section{Generation of squeezed Fock states in schemes based on particle-number measurements}
\subsection{Generation of squeezed Fock states using a two-mode Gaussian state}
In this work, we investigate the generation of squeezed Fock states (SFSs) in schemes based on particle-number measurements. An SFS is a state characterized by the following wave function:
\begin{align} \label{SFS}
\Psi_{\mathrm{\mathrm{SF}}}(x, r,n)=\frac{e^{-\frac{1}{2} e^{2 r} x^2} H_n\left(e^r x\right)}{\sqrt[4]{\pi}\sqrt{2^n n! e^{-r}}},
\end{align}
where $r$ is the squeezing parameter of SFS, and $H_n$ stands for the Hermite polynomials \cite{gradshteyn2014table}   defined such that the three  lowest of them read
\begin{equation}
\label{Her}
H_0(x) = 1, \qquad  H_1(x) = 2x,\qquad \mathrm{and}  \qquad H_2(x) = 4x^2-2.
\end{equation}
Hereafter, all wave functions are written in terms of the eigenvalues of the coordinate quadrature operator defined by the following relation
\begin{align}
 \label{x}
\hat{x}=(\hat{a}+\hat{a}^\dag)/\sqrt{2},
\end{align}
where $\hat{a}$ stands for the annihilation operator of the quantum field. 

\begin{figure} [H]
    \centering
    \includegraphics[width=0.4\linewidth]{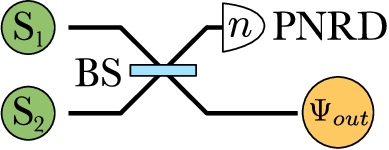}
    \caption{Scheme for generating squeezed Fock states. In the figure, $S_1$ and $S_2$  denote oscillators prepared in squeezed vacuum states, $BS$ is a beam splitter with transmission coefficient $t$, PNRD is a particle-number-resolving detector, and $\Psi_{out}$ is the generated state.}
    \label{fig:n=2}
\end{figure}

To generate such states, the work \cite{Korolev2024} presented a scheme shown in Fig. \ref{fig:n=2}.
In this scheme, the oscillators  $S_1$ and $S_2$, with squeezing parameters $r_1$ and $r_2$, respectively, are mixed on a beam splitter. This results in an entangled two-mode Gaussian state, which is described by the following wave function:
\begin{align}
\label{TMGS}
    \Psi \left(x_1,x_2 \right)=\frac{\left( a_1 a_2-b_{12}^2\right)^{1/4}}{\sqrt{\pi}}\exp \left[{-\frac{1}{2} \left(a_1 x_1^2+2 b_{12} x_1 x_2+a_2 x_2^2\right)}\right],
\end{align}
where the parameters $a_1$, $a_2$ and $b_{12}$ are related to the squeezing parameters of the oscillators and by the beam splitter transmittance \cite{Takase2021,Bashmakova_2023}. 

Next, one of the modes of the two-mode entangled state is measured using a particle-number-resolving detector (PNRD). If the scheme parameters are chosen such that the parameters of the Gaussian state satisfy the relations:
\begin{align} \label{un_sol_1}
    a_2=a_1 e^{2r}, \quad b_{12}=\sqrt{a_1^2-1}e^r,
\end{align}
then upon measuring $n$ particles, we obtain the $n$-th SFS with the squeezing parameter $r$, whose wave function coincides exactly with that in Eq. \eqref{SFS}. A notable feature of this choice is that the scheme always generates the $n$-th SFS when $n$ particles are detected. For this reason, we refer to such a choice of parameters as the universal solution, and we call any scheme whose parameters satisfy this condition a universal scheme.

As seen from Eq. (\ref{un_sol_1}), the universal solution fixes only two parameters of the input two-mode Gaussian state. The remaining parameter does not affect the generation of the $n$-th SFS, but it does affect its probability \cite{Bashmakova_2023,Korolev2024estimation}. The maximum probability of generating the $n$-th  SFS is achieved  for $a_1=2n+1$ and is equal to \cite{Korolev2024}:
\begin{align} \label{max_Prob}
    P_n=\frac{n^n}{(n+1)^{n+1}}.
\end{align}

Thus, we see that when using a two-mode Gaussian state, one can, in principle, generate SFSs with arbitrary squeezing and any value of  $n$. In practice, however, several difficulties arise. First, generating SFSs with large  $n$ requires a PNRD capable of resolving high particle numbers $n$, and the construction of such detectors remains a challenging experimental task. In addition, generating SFSs with large squeezing parameters $r$ requires states with a high degree of squeezing \cite{Korolev2024}, which is likewise difficult to achieve experimentally \cite{Vahlbruch2016}. To mitigate these limitations, one may attempt to employ schemes with multiple modes and a larger number of PNRDs. Such schemes can distribute the required input squeezing across multiple modes, as well as distribute the detected particle number $n$ across multiple outputs.

\subsection{Generation of squeezed Fock states using a multimode Gaussian state}
\begin{figure}[H]
     \centering
     \includegraphics[width=0.5\linewidth]{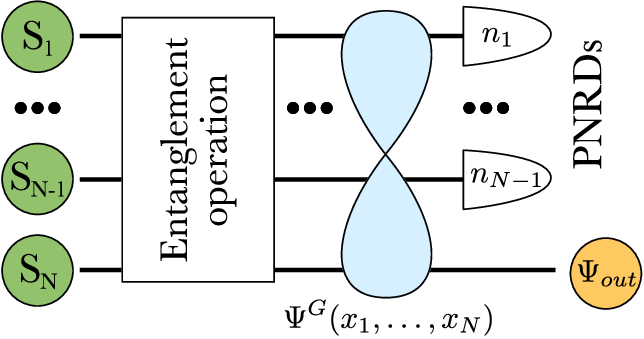}
     \caption{Scheme for generating squeezed Fock states using an $N$-mode Gaussian state and $N-1$ detectors. In the figure, $S_i$ are squeezed vacuum states, PNRDs are particle-number-resolving detectors, and $\Psi_{out}$ is the generated state.}
     \label{fig_gen_N}
 \end{figure}

Let us now use the scheme shown in Fig. \ref{fig_gen_N} for the generation.
\noindent In the presented scheme, the $N$ independent squeezed vacuum states $S_i$ are entangled using linear optical transformations. As a result, an $N$-mode Gaussian entangled quantum state is obtained, which, in the general case, is described by the following wave function:
\begin{align} \label{N_mode_gauss}
    \Psi^{G}(x_1,\dots,x_N)=\left(\frac{\det\,  \sigma}{\pi ^N}\right)^{1/4}\ e^{- \frac{1}{2}\vec{x}^T \sigma_N \vec{x}},
\end{align}
where $\vec{x}= \left(x_1, x_2, \dots, x_N\right)^T$ is a vector whose components $x_i$ are the quadrature coordinates of the $i$-th mode, and $\sigma$ is a symmetric, positive-definite matrix that fully characterizes our state:
 \begin{align}
    \sigma_N=\begin{pmatrix}
        a_{1} & b_{12} &  b_{13}& \ldots & b_{1N}\\
        b_{12} & a_{2} &  b_{23}& \ldots & b_{2N}\\
        b_{13}&b_{23}& a_{3}& \ldots& \ldots\\
        \vdots & \vdots & \vdots & \ddots & \vdots\\
        b_{1N} & b_{2N} & \dots & \dots & a_{N}
    \end{pmatrix}.
\end{align}
Next, in the resulting $N$-mode Gaussian state, $(N-1)$ modes are measured using particle-number–resolving detectors. As a result, the remaining unmeasured mode will be in the output state that can be written in the following form:
\begin{align} \label{out_int}
    \Psi_{out}(x_N)=\frac{1}{\sqrt{\mathcal{N}}}\frac{1}{\pi^{\frac{N-1}{4}}}\sqrt{\frac{2^{-n_1-\cdots -n_{N-1}}}{ n_1! \cdot \dots \cdot n_{N-1}!}}\int_{-\infty}^{\infty} dx_1\dots dx_{N-1} \Psi^{G}(x_1,\dots,x_N)e^{-\frac{x_1^2+\cdots+x_{N-1}^2}{2}}H_{n_1}(x_1) \dots H_{n_{N-1}}(x_{N-1}),
\end{align}
where $n_i$ is the number of particles measured in the $i$-th mode, and $\mathcal{N}$ is the normalization of the output state. 

Our goal is to select parameters of the multimode Gaussian state (\ref{N_mode_gauss}) such that the output state (\ref{out_int}) is the SFS (\ref{SFS}). At the same time, we want the obtained solution to be universal and independent of the order of the generated SFS. In other words, we want that, upon measuring the multimode Gaussian state with PNRDs, the resulting state is always an SFS whose order depends only on the sum of the detected particle numbers $n=n_1+\cdots +n_{N-1}$.

Having studied the states (\ref{out_int}), we found a unique universal solution that ensures that SFSs are always generated. This solution can be specified in terms of the symmetric matrix $\sigma$, which fully characterizes the Gaussian state (\ref{N_mode_gauss}):
\begin{align} \label{sol_sigma}
    \sigma_{N}=\begin{pmatrix}
        a_{1} & \sqrt{(a_1 - 1)(a_2 -1)} &  \sqrt{(a_1 - 1)(a_3 -1)}& \ldots & \sqrt{(a_1 - 1)\left(\sum \limits_{i=1}^{N-1}a_i-(N-3)\right)}e^r\\
        \dots & a_{2} &  \sqrt{(a_2 - 1)(a_3 -1)}& \ldots & \sqrt{(a_2 - 1)\left(\sum \limits_{i=1}^{N-1}a_i-(N-3)\right)}e^r\\
        \vdots & \vdots & \ddots & \dots & \vdots\\
        \dots & \dots & \dots & \dots & \left(\sum \limits_{i=1}^{N-1}a_i-(N-2)\right) e^{2r}
    \end{pmatrix},
\end{align}
where $r$ is the squeezing parameter of the generated SFS, and the parameters $a_i > 1$ are free parameters that do not affect the generated state. Substituting this matrix into the Gaussian state in Eq. (\ref{out_int}), we obtain the output state:
\begin{align}
    \Psi_{out}(x_N)= \frac{e^{r/2}}{\pi^{1/4}\sqrt{2^{n_1+\cdots+n_{N-1}}(n_1+\cdots n_{N-1})!}}\, e^{- \frac{1}{2}e^{2r} x_{N}^2}\,H_{n_1+ \dots + n_{N-1}}(e^r x_N),
\end{align}
which is precisely the squeezed Fock state defined in Eq. (\ref{SFS}). A proof of this result is presented in Appendix \ref{append_A}.

Thus, we have obtained a universal solution that always enables the generation of SFSs depending only on the sum of the detected particle numbers, $n=n_1+\cdots+n_{N-1}$. From the obtained solution (\ref{sol_sigma}), it follows that the Gaussian state still contains $N-1$ free parameters ($a_1,\, a_2,\, \dots ,\,a_{N-1}$), which do not affect the generation of the SFS. These parameters may influence the probability of generating the SFS, as well as the specific form of the physical scheme used to generate it. Let us examine this influence in detail.

\section{Probability of generation}
After obtaining the universal solution for generating SFSs, we can evaluate the probability of their generation. As we established in the previous section, the $n$-th SFS is generated when the entangled Gaussian state (\ref{N_mode_gauss}) with matrix (\ref{sol_sigma}) is measured using PNRDs. The detected particle numbers $\lbrace n_1,n_2,\dots, n_{N-1} \rbrace$ must satisfy the condition $n = n_1 + n_2 + \cdots + n_{N-1}$. This means that the probability of generating the $n$-th SFS is equal to the probability of detecting the particle numbers $\lbrace n_1,n_2,\dots, n_{N-1} \rbrace$ on the detectors under the condition $n = n_1 + n_2 + \cdots + n_{N-1}$.
Such a probability is given by the following expression:
\begin{align}
   P\left(n_1,n_2,\dots, n_{N-1}|n\right)= \frac{2({a_1 - 1})^{n_1}({a_1 - 1})^{n_2} \cdots ({a_{N-1} - 1})^{n_{N-1}}\left(n_1+n_2+\dots +n_{N-1}\right)!}{n_1!n_2!\cdots n_{N-1}! \left(\sum \limits_{i=1}^{N-1}a_i-N+3 \right)^{1+ n_1 +n_2 +\cdots + n_{N-1}}}.
\end{align}
The explicit value of the probability follows from the form of the output wave function (\ref{out_int}), the explicit evaluation of the integral (\ref{append_proof}), the form of the SFS wave function (\ref{SFS}), and the fact that for the universal solution we obtained, $\det \sigma = e^{2r}$.

When evaluating the probability in the universal scheme, it is important to consider that there are many ways to partition the number $n$ into a sum of $\lbrace n_1, n_2, \dots, n_{N-1} \rbrace$. For example, the number $4$ can be written as $1+3$, $2+2$, or $3+1$. Since the generated SFS depends only on $n$, the probability of obtaining this state in the universal scheme is given by the sum of probabilities of the form:
\begin{align}
    P(n)=\sum_{\substack{n_1,n_2,\dots,n_{N-1} \\ n_1+n_2+\cdots+n_{N-1}=n}}  P\left(n_1,n_2,\dots, n_{N-1}|n\right).
\end{align}
By summing all conditional probabilities, we obtain the following expression for the probability of generating the $n$-th SFS:
\begin{align} \label{prob_nsf}
    P(n)=2\frac{\left(\sum \limits_{k=1}^{N-1}a_k-N+1\right)^n}{\left(\sum \limits_{k=1}^{N-1}a_k-N+3\right)^{n+1}}.
\end{align} 
It follows from the obtained expression that the probability depends on the parameters of the input Gaussian state. Thus, one can optimize the generation probability of the $n$-th SFS by choosing appropriate parameters.

To simplify the analysis of the probability function (\ref{prob_nsf}), we make the following change of variables:
\begin{align} \label{un_param}
    X=\sum \limits_{k=1}^{N-1}a_k-N+2.
\end{align}
Considering the conditions $a_i>1$, it is straightforward to verify that $X>1$. Hereafter, we refer to the parameter $X$ as the universal parameter of the scheme. With this substitution, the resulting probability becomes a function of a single parameter:
\begin{align} 
    P(n)=2\frac{\left(X-1\right)^n}{\left(X+1\right)^{n+1}}.
\end{align} 
In this case, the probability of generating the $n$-th SFS does not explicitly depend on the number of input modes $N$. For any two numbers of input modes $N_1$ and $N_2$, one can choose the scheme parameters $a_i$ and $a_{i'}$ such that the equality $\sum \limits_{k=1}^{N_1-1}a_k-N_1+2=\sum \limits_{k=1}^{N_2-1}a_k'-N_2+2$. 

To maximize the probability of generating the $n$-th SFS, we must choose the scheme parameters such that $X = 2n + 1$. In this case, we obtain the probability
\begin{align} 
    P(n)=\frac{n^n}{(n+1)^{n+1}}.
\end{align} 
This probability exactly coincides with the value of the maximal probability (\ref{max_Prob}) obtained for the case of a two-mode Gaussian state. This further demonstrates the independence of the probability of generating the $n$-th SFS from the number of input modes. The maximal probability is determined solely by the order $n$ of the generated SFS.

The dependence of the SFS generation probability on the sum of all free parameters $X$ implies that, after maximizing the probability, there remain $N-2$ free parameters in the scheme. Let us assess their influence on the SFS generation scheme.

\section{Scheme for generating squeezed Fock states using a multimode Gaussian state}
We now describe how to experimentally generate the multimode Gaussian states required for producing SFSs.

As noted earlier, the matrix (\ref{sol_sigma}) establishes a relation between the parameters of the generation scheme and those of the input states. Such a symmetric positive-definite matrix can be represented as follows:
\begin{align} \label{decomp_sigm}
    \sigma_N=\mathcal{O} \mathcal{D}\mathcal{O}^{T}, 
\end{align}
where $\mathcal{O}$ is an orthogonal matrix and $\mathcal{D}$ is a diagonal matrix. Both matrices $\mathcal{O}$ and $\mathcal{D}$ are real. From a physical point of view, this means that to generate states described by the matrix $\sigma_N$, it is sufficient to consider only the mixing of squeezed states in orthogonal quadratures using beam splitters 
\cite{Takase2021,Bashmakova_2023,Rendell2005}. 

\begin{figure}[H]
    \centering
    \includegraphics[width=0.4\linewidth]{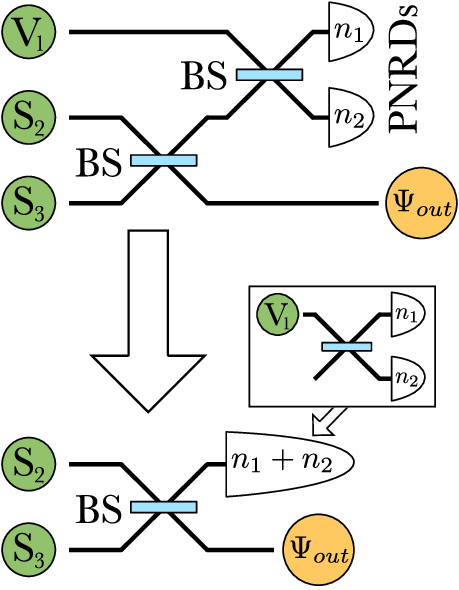}
    \caption{Scheme for generating a squeezed Fock state using a three-mode Gaussian state. In the figure, $S_2$ and $S_3$ are quantum oscillators in squeezed vacuum states, $V_1$ is a quantum oscillator in the vacuum state, BS denotes beam splitters, and $\Psi_{out}$ is the output state.}
    \label{fig_3mode_equ}
\end{figure}

To obtain a decomposition of the matrix (\ref{decomp_sigm}) in terms of beam splitters, let us introduce the beam splitter matrix $BS^{kl}(t)$, which mixes the $k$-th and $l$-th modes with transmission coefficient $t$. Using such matrices, we can write the following decomposition of the matrix $\sigma_N$:
\begin{align}
    \sigma_N=\left(BS^{1,2}(t_{1})\cdots BS^{N-2,N-1}(t_{N-2})BS^{N-1,N}(t_{N-1}) \right)\mathcal{D} \left(BS^{1,2}(t_{1})\cdots BS^{N-2,N-1}(t_{N-2})BS^{N-1,N}(t_{N-1})\right)^T.
\end{align}
The transmission coefficients of beam splitters are as follows:
\begin{align}
    &t_{1}=\sqrt{\frac{a_2-1}{a_1+a_2-2}},\\
    &t_{2}=\sqrt{\frac{a_3-1}{a_1+a_2+a_3-3}},\\
    &\dots,\\
    &t_{N-2}=\sqrt{\frac{a_{N-1}-1}{\sum \limits_{i=1}^{N-1} a_i-N+1}},\\
    &t_{N-1}=\sqrt{\frac{1}{2}\left(1+\frac{\left(\sum \limits_{i=1}^{N-1} a_i-N+2\right)\left(1-e^{2r}\right)}{\sqrt{\left(\sum \limits_{i=1}^{N-1} a_i-N+2\right)^2(1+e^{2r})^2-4e^{2r}}}\right)}. \label{t_N}
\end{align}
The diagonal matrix has the following form:
\begin{align}
\mathcal{D}=\text{diag}\left(1,1,\cdots,1,e^{2r_{N-1}},e^{2r_N}\right),
\end{align}
where 
\begin{align}
   e^{2r_{N-1}}=\frac{1}{2}\left(\left(\sum_{k=1}^{N-1}a_k-N+2\right)\left(1+e^{2r}\right)+\sqrt{\left(\sum_{k=1}^{N-1}a_k-N+2\right)^2\left(1+e^{2r}\right)^2-4e^{2r}}\right), \label{sq_1}\\
   e^{2r_N}=\frac{1}{2}\left(\left(\sum_{k=1}^{N-1}a_k-N+2\right)\left(1+e^{2r}\right)-\sqrt{\left(\sum_{k=1}^{N-1}a_k-N+2\right)^2\left(1+e^{2r}\right)^2-4e^{2r}}\right). \label{sq_2}
\end{align}
From the obtained decomposition, it follows that, to generate the SFS in a scheme with an $N$-mode Gaussian state, it is necessary to use two squeezed vacuum states and $N-2$ vacuum states. The resulting SFS generation scheme for the three-mode case is shown in Fig. \ref{fig_3mode_equ}.
\noindent It is also important to note that the presented decomposition into beam splitter transformations is not unique. Other decompositions of the orthogonal matrix $\mathcal{O}$ can be constructed; however, the diagonal matrix $\mathcal{D}$ does not depend on the particular decomposition. In all cases, only two squeezed vacuum states are used in the generation scheme.

By analyzing the expressions for the oscillator squeezing parameters (\ref{sq_1}) and (\ref{sq_2}), as well as the expression for the beam splitter transmittance (\ref{t_N}), one can see that a universal parameter (\ref{un_param}) is distinguished in them. This means that they do not depend on the number of input oscillators, but only on the squeezing of the generated SFS. The remaining $N-2$ free parameters $a_i$ affect only the transmittances $t_1$, ..., $t_{N-2}$. In other words, all transmission coefficients except $t_{N-1}$ are free parameters and can be chosen arbitrarily.

Thus, the problem of generating SFSs in the multimode case can be completely reduced to the two-mode problem (see Fig. \ref{fig_3mode_equ}). In both the two-mode and multimode cases, we can generate the SFS $\hat{S}(r)\lvert n\rangle$ with identical probabilities by using two squeezed oscillators with identical squeezing parameters. The only difference lies in the detection stage. In the two-mode case, one must detect $n$ particles with a single detector, whereas in the multimode case, the $n$ particles can be distributed among $N-1$ detectors. Such a distribution is arbitrary.

The obtained result is of practical interest. At present, existing PNRDs are limited in the number of particles they can resolve. By using several such PNRDs, however, one can generate SFSs with larger values of $n$.

\section{Robustness to detection errors}
When discussing detectors, it is essential to keep in mind that ideal detectors do not exist. Real detectors are characterized by a detection efficiency parameter $\eta$, which is always less than unity. The presence of detector imperfections leads to a degradation of the SFS generation process \cite{Bashmakova_2024}. The purpose of this section is to quantify the degradation of the SFS generated in a scheme with $N$ inputs as a function of the efficiencies of the $N-1$ detectors.

\begin{figure} [H]
    \centering
    \includegraphics[width=0.5\linewidth]{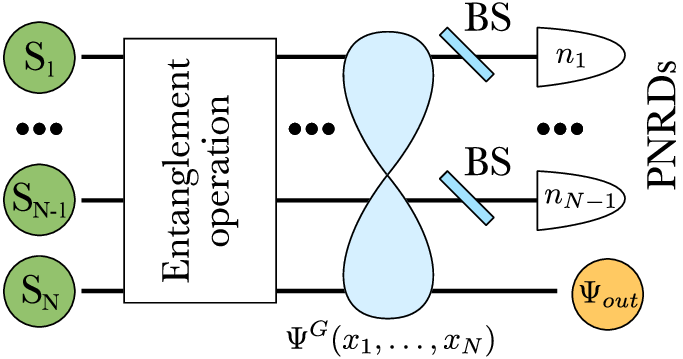}
    \caption{Scheme for generating a squeezed Fock state using non-ideal detectors. In the figure, $S_i$ denotes the squeezed vacuum state, PNRDs are particle-number–resolving detectors, BS denotes a beam splitter with transmittance $t=\eta$, and $\Psi_{out}$ is the generated state.}
    \label{fig_eff_det}
\end{figure}

Let the efficiencies of all detectors be identical and equal to $\eta <1$. Mathematically, the effect of detectors with efficiency $\eta$ on the generated state can be described as a combination of beam splitters with transmission coefficient $t = \eta$ and ideal PNRDs \cite{Kilmer2019}. The additional beam splitters mix vacuum states into the scheme, which are measured together with the signal states by the detectors. A scheme with non-ideal detectors is shown in Fig. \ref{fig_eff_det}.
At the output of the presented scheme, the state is described by the following wave function:
\begin{multline} 
    \tilde{\Psi}_{out}(x_N,x_1^v,\dots,x_{N-1}^v)=\frac{1}{\sqrt{\tilde{\mathcal{N}}}}\int_{-\infty}^{\infty} dx_1\cdots dx_{N-1} \Psi^{G}(\eta x_1+\sqrt{1-\eta^2}x^v_1,\eta x_2+\sqrt{1-\eta^2}x^v_2,\dots,x_N)\\
    \times \Psi_{\text{vac}}(\sqrt{1-\eta^2}x_1-\eta x^v_1)\Psi_{\text{vac}}(\sqrt{1-\eta ^2}x_2-\eta x^v_2)\cdots \Psi_{n_1}(x_1)\cdots \Psi_{n_{N-1}}(x_{N-1}),
\end{multline}
where $\Psi^{G}(x_1,\dots,x_N)$ is the wave function of the input multimode Gaussian state, $\Psi_{n}(x)$ is the wave function of the Fock state, $\Psi_{\text{vac}}(x)$ is the wave function of the vacuum state, $x_i^v$ are the generalized coordinates of the vacuum state mixed with the $i$-th quantum oscillator, and $\tilde{\mathcal{N}}$ is the normalization factor of the state.

To quantitatively assess how close the generated state is to the SFS with squeezing parameter $r$, we use the fidelity, which in this case is given by:
\begin{align}  \label{fidelity_ffin}
 F(\eta)=\int dx_{N}\left|\int dx_{1}^v\cdots dx^v_{N-1}\tilde{\Psi}_{out}(x_N,x_1^v,\dots,x_{N-1}^v)\Psi_{\mathrm{\mathrm{SF}}}(x_N,r, n)\right|^{2},
 \end{align}
where $n_1+\cdots+n_{N-1}=n$ is the total number of particles detected in the scheme. 
 
For the wave function of the multimode Gaussian state with the matrix $\sigma$ defined by Eq. (\ref{sol_sigma}), the value of the fidelity (\ref{fidelity_ffin}) can be calculated analytically. The fidelity takes the following form:
\begin{align}
    F(\eta)=  \left(\frac{\left(\sum \limits_{k=1}^{N-1} a_k-N+1\right)\eta^2+2}{\sum \limits_{k=1}^{N-1} a_k-N+3}\right)^{n+1}.
\end{align}
It follows from the obtained expression that the fidelity does not depend on the squeezing parameter of the SFS, but only on the universal parameter $X$ defined by Eq. (\ref{un_param}):
\begin{align} \label{fid}
    F_{n}(\eta)=  \left(\frac{(X-1)\eta ^2+2}{X+1}\right)^{n+1}.
\end{align}
This means that the fidelity also does not depend on the number of non-ideal detectors used. That is, adding additional non-ideal detectors to the scheme does not degrade the generated state any more than using a single such detector.

Furthermore, by analyzing Eq. (\ref{fid}), we see that as $X$ increases ($X>1$), the fidelity decreases for any values of $\eta$ and $n$. This implies that, to achieve higher fidelity, it is preferable to choose a smaller universal parameter. For small values of $X$, the required oscillator squeezing [see Eqs. (\ref{sq_1}) and (\ref{sq_2})] is also reduced, but so does the probability of generating the SFS (\ref{prob_nsf}). Thus, by sacrificing the generation probability of the SFS, one can relax the squeezing requirements of the oscillators used and increase the fidelity of the generated state.

\section{Comparison of universal and non-universal generation schemes}
In the previous sections, we analyzed the universal scheme for generating SFSs. We found that such a scheme does not provide any advantage in terms of the required squeezing of input oscillators: the squeezing does not depend on the number of modes. To obtain an advantage nonetheless, we now consider a non-universal scheme. In this scheme, the parameters are tuned such that an SFS is generated only when a specific sequence of particle numbers $n_1, n_2, \ldots, n_{N-1}$ is detected. Otherwise, a different quantum state is produced.

\begin{figure} [H]
    \centering
    \includegraphics[width=0.5\linewidth]{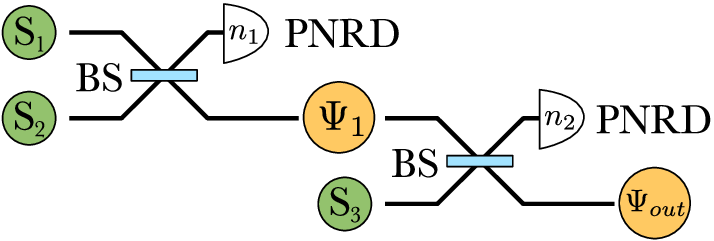}
    \caption{Cascade scheme for generating squeezed Fock states. In the figure, $S_1$, $S_2$, and $S_3$ denote squeezed vacuum states, $\Psi_1$ is the state at the output of the first stage of the scheme, and $\Psi_{out}$ is the final output state.}
    \label{fig_casc}
\end{figure}

As a non-universal scheme, we consider the so-called cascade scheme, in which particle measurements are performed sequentially. A special case of this scheme for three modes is shown in Fig. \ref{fig_casc}.
\noindent In the scheme, two orthogonally squeezed vacuum states $S_1$ and $S_2$ (with squeezing parameters $r_1$ and $r_2$) are mixed on a beam splitter with transmittance $t_1$. One of the modes is then measured using a PNRD (measurement outcome $n_1$). The unmeasured mode enters the second stage of the scheme, where it is mixed with another squeezed vacuum state $S_3$ (with squeezing parameter $r_3$). Subsequently, one of the remaining modes is measured with another PNRD (measurement outcome $n_2$). The remaining unmeasured mode is in the output state.

The parameters of the scheme shown in Fig. \ref{fig_casc} can be chosen such that, similarly to the universal scheme (Fig. \ref{fig_3mode_equ}), an SFS is generated at the output. Let us compare these two schemes. The comparison will be performed with respect to two criteria: the probability of SFS generation and the resource requirements required for their implementation. We consider two specific cases corresponding to SFS generation with $n=2$ and $n=3$.

In this comparison, we exclude events in which one of the detectors in the cascade scheme registers zero particles, since such events are trivial and effectively reduce the scheme to a two-mode universal one. This is because detecting zero particles does not introduce non-Gaussianity into the scheme. For simplicity, we further assume that all detectors are ideal and operate with unit efficiency.

To generate the SFS state $\hat{S}(r)\lvert 2\rangle$ in the cascade scheme, one must detect a single particle at each of the two detectors. However, in this case, the generated state cannot be reduced to the SFS by an appropriate choice of the scheme parameters. In other words, it is impossible to find a cascade scheme in which detecting one particle at each of the two detectors results in the generation of the SFS. The maximum fidelity between the generated state and the state $\hat{S}(r)\lvert 2\rangle$ is given by the following expression:
\begin{align}
   F= \frac{2}{3}+\frac{4}{3-9 \cosh \left(2 \left(r-r_3\right)\right)}.
\end{align}
The above expression shows that the fidelity is limited from above by $2/3$, indicating that the generated state differs significantly from the SFS.

\begin{figure}[H]
    \centering
    \includegraphics[width=0.5\linewidth]{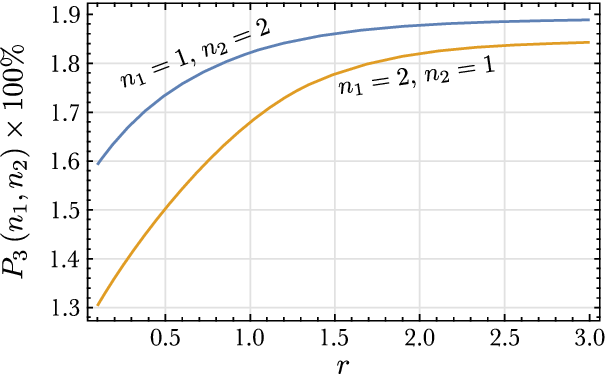}
    \caption{Probability of generating squeezed Fock states in a cascaded scheme as a function of the squeezing parameter $r$. Different colors correspond to the two cases: blue denotes the case for $n_1=1$ and $n_2=2$, , while orange denotes the case for $n_1=2$ and $n_2=1$.}
    \label{fig:prob_casc}
\end{figure}

Let us now compare the generation of the SFS $\hat{S}(r)\lvert 3\rangle$ in the two schemes. In the cascade scheme, generating the state under consideration requires detecting two particles on one detector and one particle on the other. Depending on the order of these detection events ($n_1 = 1$, $n_2 = 2$ or $n_1 = 2$, $n_2 = 1$), the generation scheme itself changes. In other words, a scheme configured to generate an SFS upon measuring $n_1 = 1$ and $n_2 = 2$ will not produce it when the outcomes are $n_1 = 2$ and $n_2 = 1$. This is in sharp contrast to the universal scheme, where the order of detection is irrelevant and only the total number $n_1 + n_2$ matters. This difference directly affects the generation probability: only one specific measurement outcome leads to the desired SFS. Fig. \ref{fig:prob_casc} shows the maximal achievable probabilities of SFS generation $P_3(n_1, n_2)$ for different squeezing parameters $r$ in cascade schemes.
\noindent From the plot, it is clear that in the cascade scheme the generation probabilities depend on the squeezing parameter $r$ of the generated SFS. In the universal scheme, the probability does not depend on the squeezing and is approximately $10.5 \%$. Thus, we observe a significant advantage in the generation probability for the universal scheme. Moreover, as the total number of detected particles  $n_1+n_2$ increases, the difference between the SFS generation probabilities in the universal and cascade schemes becomes increasingly pronounced. This is related to the number of measurement outcomes that lead to SFS generation: in the cascade scheme, there is always only one such outcome, whereas in the universal scheme, the number of favorable outcomes grows with increasing $n_1+n_2$.

\begin{figure}[H]
    \centering
    \includegraphics[width=0.5\linewidth]{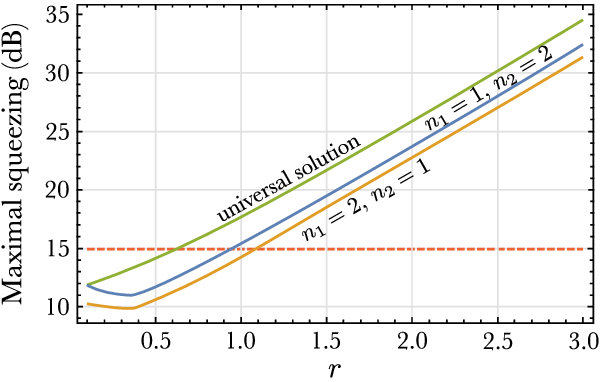}
    \caption{The squeezing degree of the most strongly squeezed oscillators required to generate an SFS with a given squeezing parameter $r$. Different colors correspond to different cases: blue denotes the maximal oscillator squeezing in the cascade scheme for $n_1=1$ and $n_2=2$; orange denotes the maximal oscillator squeezing in the cascade scheme for $n_1=2$ and $n_2=1$; green denotes the maximal oscillator squeezing in the universal scheme; and the red dashed line indicates the 15 dB level. The squeezing is measured in dB.}
    \label{fig:max_sq}
\end{figure}

Let us now estimate the physical resources required for SFS generation in the universal and cascade schemes. In both schemes, squeezed vacuum states are used and are entangled via beam splitters. Therefore, the primary physical resource is the squeezing of the input oscillators. This resource, however, is severely limited. The record level of squeezing demonstrated experimentally to date is $15$ dB. Fig. \ref{fig:max_sq} shows the squeezing levels of the most strongly squeezed oscillators required to generate an SFS with a given squeezing parameter $r$ and with the maximal achievable probability in the universal and cascade schemes. 
\noindent  From the plot, it is evident that the maximal squeezing required for the input oscillators in the universal scheme is higher than in the cascade schemes. This is because the cascade scheme employs three squeezed oscillators, rather than two as in the universal scheme, effectively distributing the total squeezing resource over three inputs. As a result, for the same degree of input squeezing, the cascade scheme enables the generation of SFSs with a larger squeezing parameter.

\begin{figure}[H]
    \centering
    \includegraphics[width=0.5\linewidth]{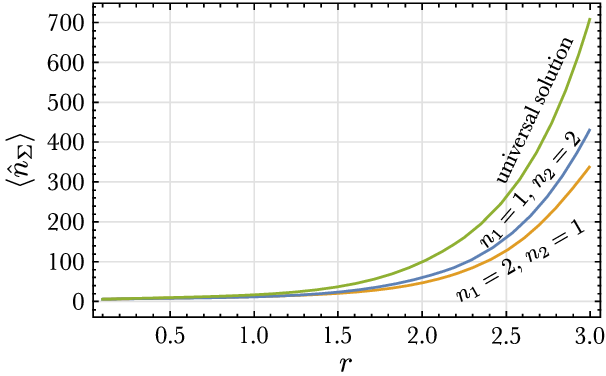}
    \caption{Total average particle number in the oscillators required to generate an SFS with squeezing parameter $r$ in different schemes. Different colors correspond to three cases: blue denotes the cascade scheme with $n_1 = 1$ and $n_2 = 2$; orange denotes the cascade scheme with $n_1 = 2$ and $n_2 = 1$; and green denotes the universal scheme.}
    \label{fig:energy}
\end{figure}

We can also assess the required resources from an energetic perspective, namely, how much energy is needed to prepare the input resources for SFS generation in different schemes. Since the energy is proportional to the number of particles in the states, we will use the total average particle number in the squeezed oscillators employed for SFS generation as an energy measure: $\langle \hat{n} _{\Sigma}\rangle=\langle\hat{n}_1\rangle+\langle\hat{n}_2\rangle+\langle\hat{n}_3\rangle$. Fig. \ref{fig:energy} shows the dependence of the total average number of particles in the oscillators, $\langle \hat{n}_{\Sigma} \rangle$, on the squeezing parameter $r$ of the generated SFS.
\noindent From the plot, it is evident that the highest energy is required in the universal generation scheme. The lowest energy is achieved in the cascade scheme when $n_1 = 2$ and $n_2 = 1$. By comparing Fig. \ref{fig:energy} and Fig. \ref{fig:prob_casc}, one can observe a correlation between the required energy and the generation probabilities. The higher the maximum probability of generating an SFS, the more energy the scheme requires.

\section{Conclusion}
In this work, we conducted a comprehensive study of the scheme for generating squeezed Fock states (SFSs) via particle-number measurements in the modes of a multimode Gaussian state. We identified a form of the Gaussian wave function that enables SFS generation for any outcome of a particle-number measurement. We then developed a generation scheme for this Gaussian state. It turns out that the resulting scheme is fully equivalent to a two-mode scheme, in which $n$ particles in the measured channel are distributed across multiple detectors. This scheme does not reduce the squeezing requirements of the input oscillators, but it allows the detection of a larger total number of particles on detectors with limited resolving capability.

We obtained analytical expressions for the SFS generation probabilities and the required squeezing degree of the oscillators. Additionally, we derived an expression for the fidelity of the generated SFSs when non-ideal detectors are used. Analyzing these expressions, we concluded that one can sacrifice the SFS generation probability to reduce the required squeezing of the input oscillators and increase the fidelity of the generated SFS.

We compared the universal scheme with a non-universal generation scheme, in which the scheme parameters must be adjusted for specific measured particle numbers to generate an SFS. We found that the non-universal scheme provides an advantage in terms of reduced oscillator squeezing but performs significantly worse than the universal scheme in terms of the SFS generation probability.

The obtained results highlight the importance of choosing the generation scheme depending on the specific task, whether the requirement is universality or the need to minimize energy costs.

\vspace{0.5 cm}

This research was supported by the Theoretical Physics and Mathematics Advancement Foundation "BASIS"\, (Grant No. 24-1-3-14-1).

\bibliography{bibliography}

\appendix
\section{Proof of the universality of the solution} \label{append_A}
To prove the existence of a universal solution that allows us to choose the parameters of the $N$-mode Gaussian state so that when we measure $N-1$ of its modes using particle number detectors we always obtain a squeezed Fock state, we must prove that the following equality holds for the integral:
\begin{multline} \label{append_proof}
I_{\lbrace n_1,n_2,\dots, n_{N-1}\rbrace}=\int_{-\infty}^{\infty} dx_1...dx_{N-1} e^{- \frac{1}{2}\vec{x}^T \sigma_N \vec{x}}e^{-\frac{x_1^2+\cdots+x_{N-1}^2}{2}}H_{n_1}(x_1) \dots H_{n_{N-1}}(x_{N-1})=\\
   =(-1)^{m_1+m_2+\dots+m_{N-1}}\sqrt{\frac{2 \pi^{N-1} (a_1-1)^{{m_1}}\cdots (a_{N-1}-1)^{{m_{N-1}}}}{\left(\sum \limits_{k=1}^{N-1} a_k-N+3\right)^{{m_1+m_2+\dots+m_{N-1}+1}}}} e^{- \frac{e^{2r} x_{N}^2}{2}}H_{m_1+ \dots + m_{N-1}}(e^r x_N),
\end{multline}
where $\vec{x}= \left(x_1, x_2, \dots, x_N\right)^T$ is a vector whose components $x_i$ are the quadrature coordinates of the $i$-th mode, and $\sigma$ is a symmetric, positive-definite matrix that fully characterizes our state:
\begin{align}
    \sigma_N=\begin{pmatrix}
        a_{1} & b_{12} &  b_{13}& \ldots & b_{1N}\\
        b_{12} & a_{2} &  b_{23}& \ldots & b_{2N}\\
        b_{13}&b_{23}& a_{3}& \ldots& \ldots\\
        \vdots & \vdots & \vdots & \ddots & \vdots\\
        b_{1N} & b_{2N} & \dots & \dots & a_{N}
    \end{pmatrix}.
\end{align}

To this end, in the matrix $\sigma_N$ we relate the off-diagonal element $b_{ij}$ to the diagonal elements $a_i$ as follows:
\begin{align}
    b_{ij}=\sqrt{(a_{i}-1)(a_{j}-1)}, \qquad  i,j\neq N \\
    b_{iN}=\sqrt{(a_{i} - 1)\left(\sum \limits_{k=1}^{N-1}a_{k}-N+3)\right)}e^r, \qquad  i\neq N
\end{align}
and we define the element $a_{N}$ as:
\begin{align}
    a_{N}=\left(\sum \limits_{k=1}^{N-1}a_{k}-N+2\right) e^{2r}.
\end{align}
In addition, to complete the proof, we employ the integral representation of the Hermite polynomial:
\begin{align} \label{apend_int_herm}
    H_m(x_n)=\frac{2^m (-i)^m e^{x_n^2}}{\sqrt{\pi}}\int_{-\infty}^{\infty}dt_n e^{-t_n^2 + 2 t_n x_n i} t_n^m.
\end{align}

Substituting the elements of the matrix $\sigma _N$ into the integral representation of the Hermite polynomial, the left-hand side of Eq. (\ref{append_proof}) can be rewritten as follows:
\begin{multline} \label{append_integr}
    I_{\lbrace n_1,n_2,\dots, n_{N-1}\rbrace}=\int_{-\infty}^{\infty}dt_1...dt_{N-1}\frac{2^{m_1+\ldots+m_{N-1}} (-i)^{m_1+\dots+m_{N-1}} }{(\sqrt{\pi})^{N-1}}e^{-t_1^2 - \dots - t_{N-1}^2} t_1^{m_1} \dots t_{N-1}^{m_{N-1}} \\
    \times \int_{-\infty}^{\infty} dx_1 \dots dx_{N-1}  e^{-\sqrt{(a_1 - 1)(a_2 -1)}x_1 x_2 -  \sqrt{(a_1 - 1)(a_3 -1)}x_1 x_3 - \dots - \sqrt{(a_1 - 1)\left(\sum \limits_{k=1}^{N-1}a_k-(N-3) \right)}e^r x_1 x_N }   \\
   \times  e^{-\sqrt{(a_2 - 1)(a_3 -1)}x_2 x_3 -  \sqrt{(a_2 - 1)(a_4 -1)}x_2 x_4 - \dots - \sqrt{(a_2 - 1)\left(\sum \limits_{k=1}^{N-1}a_k-(N-3) \right)}e^r x_2 x_N }\\ 
   \times \dots \\
   \times e^{-\sqrt{(a_{N-1} - 1)\left(\sum \limits_{k=1}^{N-1}a_k-(N-3) \right)}e^r x_{N-1} x_N }\\
     \times e^{x_1^2 + \dots + x_{N-1}^2} e^{-\frac{a_1}{2}x_1^2-\frac{a_2}{2}x_2^2-\dots-\frac{a_{N-1}}{2}x_{N-1}^2-\frac{\left(\sum \limits_{k=1}^{N-1}a_k-(N-2) \right) }{2}e^{2r}x_{N}^2}  \\
    \times e^{-\frac{x_1^2}{2}-\frac{x_2^2}{2}-\dots-\frac{x_{N-1}^2}{2}} e^{2t_1 x_1 i + 2 t_2 x_2 i + \dots + 2 t_{N-1} x_{N-1} i}.
\end{multline}
Let us now take the integral over $x_1$ separately:
 \begin{multline} 
 \int_{-\infty}^{\infty} dx_1 e^{-\frac{a_1-1}{2}x_1^2} e^{- x_1 \left(\sqrt{(a_1 - 1)(a_2 -1)} x_2 + \sqrt{(a_1 - 1)(a_3 -1)}x_3 + \dots + \sqrt{(a_1 - 1)\left(\sum \limits_{k=1}^{N-1}a_k-(N-3) \right)}e^r x_N -2 t_1 i\right)} \\
 =\sqrt{\frac{2 \pi}{a_1 - 1}}  e^{\frac{\left(\sqrt{(a_1 - 1)(a_2 -1)} x_2 +  \sqrt{(a_1 - 1)(a_3 -1)}x_3 + \dots + \sqrt{(a_1 - 1)\left(\sum \limits_{k=1}^{N-1}a_k-(N-3) \right)}e^r x_N -2 t_1 i\right)^2}{2(a_1 - 1)}}.
 \end{multline}
Substituting the obtained value of the integral into Eq. (\ref{append_integr}) and expanding the brackets, we find that all terms quadratic in  $x_i$ (except for $x_N$) and all cross terms $x_ix_j$ in the exponent cancel out. As a result, the integral of interest is transformed as follows:
 \begin{multline}  \label{append_2_step}
    I_{\lbrace n_1,n_2,\dots, n_{N-1}\rbrace}=\int_{-\infty}^{\infty}dt_1...dt_{N-1}\frac{2^{m_1+\ldots+m_{N-1}} (-i)^{m_1+\dots+m_{N-1}} }{(\sqrt{\pi})^{N-1}}e^{-t_1^2 - \dots - t_{N-1}^2} t_1^{m_1} \dots t_{N-1}^{m_{N-1}} \\
\times e^{- \frac{2 t_1^2}{(a_1 - 1)}} e^{\frac{e^{2r} x_{N}^2}{2}} e^{-2\sqrt{\frac{\left(\sum \limits_{i=1}^{N-1}a_i-(N-3) \right)}{(a_1 - 1)}}e^r t_1 x_N i} \sqrt{\frac{2 \pi}{a_1 - 1}}\\
 \times \int_{-\infty}^{\infty}dx_2...dx_{N-1} e^{2 x_2 i \left(t_2 - t_1 \sqrt{\frac{a_2 -1}{a_1 - 1}}\right) + \dots +2 x_{N-1} i \left(t_{N-1} - t_1 \sqrt{\frac{a_{N-1} -1}{a_1 - 1}}\right)}
\end{multline}
Taking into account that
 \begin{align}
 \int_{-\infty}^{\infty} dx_k e^{2 x_k i \left(t_k - t_1 \sqrt{\frac{a_k -1}{a_1 - 1}}\right)}=\pi  \delta \left(t_k - t_1 \sqrt{\frac{a_k -1}{a_1 - 1}}\right),
 \end{align}
 Eq. (\ref{append_2_step}) can be written as follows:
  \begin{multline}  \label{append_3_step}
    I_{\lbrace n_1,n_2,\dots, n_{N-1}\rbrace}=\int_{-\infty}^{\infty}dt_1\dots dt_{N-1}\frac{2^{m_1+\ldots+m_{N-1}} (-i)^{m_1+\dots+m_{N-1}} }{(\sqrt{\pi})^{N-1}}e^{-t_1^2 - \dots - t_{N-1}^2} t_1^{m_1} \dots t_{N-1}^{m_{N-1}} \\
\times e^{- \frac{2 t_1^2}{(a_1 - 1)}} e^{\frac{e^{2r} x_{N}^2}{2}} e^{-2\sqrt{\frac{\left(\sum \limits_{k=1}^{N-1}a_k-N+3 \right)}{(a_1 - 1)}}e^r t_1 x_N i} \sqrt{\frac{2 \pi}{a_1 - 1}}\\
 \times \pi^{N-2}  \delta \left(t_2 - t_1 \sqrt{\frac{a_2 -1}{a_1 - 1}}\right) \dots  \delta\left(t_{N-1} - t_1 \sqrt{\frac{a_{N-1} -1}{a_1 - 1}}\right).
\end{multline}
It is then straightforward to perform the integrals over $t_2, t_3,\dots, t_{N-1}$. As a result, we obtain the following expression:
\begin{multline}  \label{append_4_step}
    I_{\lbrace n_1,n_2,\dots, n_{N-1}\rbrace}  =\sqrt{\frac{2 \pi ^{N-2}}{a_1 - 1}} \left(\frac{a_2-1}{a_1-1}\right)^{\frac{m_2}{2}}\cdots \left(\frac{a_{N-1}-1}{a_1-1}\right)^{\frac{m_{N-1}}{2}} \int_{-\infty}^{\infty}dt_1 2^{m_1+\ldots+m_{N-1}} (-i)^{m_1+\dots+m_{N-1}}\\
 \times   e^{- \frac{ \left(\sum_{k=1}^{N-1} a_k-N+3\right) t_1^2}{(a_1 - 1)}-2\sqrt{\frac{\left(\sum \limits_{k=1}^{N-1}a_k-N+3\right)}{(a_1 - 1)}}e^r t_1 x_N i}  e^{\frac{e^{2r} x_{N}^2}{2}} t^{m_1+m_2+\dots +m_{N-1}}.
\end{multline}
Using the integral representation of the Hermite polynomial (see Eq. (\ref{apend_int_herm})), the final expression for the integral in Eq. (\ref{append_proof}) is obtained:
\begin{multline}  \label{append_5_step}
    I_{\lbrace n_1,n_2,\dots, n_{N-1}\rbrace} =\sqrt{\frac{2 \pi^{N-1} (a_1-1)^{{m_1}}\cdots (a_{N-1}-1)^{{m_{N-1}}}}{\left(\sum \limits_{k=1}^{N-1} a_k-N+3\right)^{{m_1+m_2+\dots+m_{N-1}+1}}}}e^{-\frac{e^{2r} x_{N}^2}{2}} H_{m_1+m_2+\dots +m_{N-1}}(e^{r} x_N)\\
    \times (-1)^{m_1+m_2+\dots+m_{N-1}}.
\end{multline}

\end{document}